\documentclass[12pt,english,aps, final]{revtex4}
\usepackage[T1]{fontenc}
\usepackage[latin9]{inputenc}
\usepackage{amsmath}
\usepackage{amssymb}
\usepackage{stmaryrd}
\usepackage{graphicx}
\usepackage{esint}

\makeatletter
\@ifundefined{textcolor}{}
{%
 \definecolor{BLACK}{gray}{0}
 \definecolor{WHITE}{gray}{1}
 \definecolor{RED}{rgb}{1,0,0}
 \definecolor{GREEN}{rgb}{0,1,0}
 \definecolor{BLUE}{rgb}{0,0,1}
 \definecolor{CYAN}{cmyk}{1,0,0,0}
 \definecolor{MAGENTA}{cmyk}{0,1,0,0}
 \definecolor{YELLOW}{cmyk}{0,0,1,0}
}

\makeatother

\usepackage{babel}
\begin{document}

\title{Crossed Topology in Two-Loop Dispersive Approach}

\author{A. Aleksejevs}

\affiliation{Grenfell Campus of Memorial University, Canada}
\begin{abstract}
We extend existing dispersive approach in subloop insertion to the
case of crossed two-loop box type topologies. Based on the ideas of
the Feynman trick, mass shift approach and dispersive representation
of two-point Passarino-Veltman function we expressed two-loop scalar
diagrams in the compact analytical form suitable for the automatization
of the calculations. The results are expressed in a way that the numerical
integration over Feynman and dispersive parameters and differentiation
with respect to mass shift parameters are required in the final stage
only.
\end{abstract}
\maketitle

\section{Introduction}

The experimental searches for the physics beyond the standard model
such as MOLLER \cite{key-2} frequently require calculations of the
observables to a high degree of precision. This can be achieved by
accounting for the next-to-next to leading order (NNLO) perturbative
contributions in the scattering matrix element. That translates to
the evaluations of two-loop Feynman diagrams, but in general this
is not a trivial task. Particularly, a full set of two-loop electroweak
corrections is close to impossible task to complete without some sort
of automatization, and tremendous effort is already invested in the
development of the various approaches for evaluation of the two-loop
diagrams. Development of the techniques in the two-loop self-energies
and vertex functions calculations is outlined in \cite{key-1,key-13,key-3,key-4}
and has been extended in more recent work of \cite{key-5,key-6,key-7,key-8,key-9,key-10}.
Two-loop n-point integrals have been evaluated in \cite{key-11,key-12}
using techniques of sector decomposition. In electroweak physics,
authors in \cite{key-14} studied two-loop fermionic contributions
to the effective Weinberg mixing angle. Two-loop electroweak corrections
to the $M_{W}-M_{Z}$ mass correlation was studied in \cite{key-15}.
In \cite{key-17,key-19,key-20,key-21}, the effort was directed to
the studies of dominant contributions of the two-loop electroweak
corrections to the parity-violating asymmetry in Moller scattering.
In \cite{AA1}, a general approach was developed to deal with the
two-loop diagrams calculations in the case of arbitrary tensor structure,
which employed an idea of dispersive subloop insertion. A general
notion outlined in \cite{AA1} is that for many two-loop topologies
it is possible to join all but one propagators in the sub-loop insertion
in a way that integration momenta of the second loop is not present
when applying Feynman trick. This way, it was possible to reduce subloop
insertion to two-point Passarino-Veltman function and later replace
subloop insertion by the effective propagator using dispersive representation.
After that, the second loop integration can be carried out analytically
in the Passarino-Veltman basis. In this paper, we extend our approach
used in \cite{AA1} to the cases where subloop insertion has crossed-box
type of topology. For this topology, it is not possible to join all
propagators except one, as we did before, and hence it would be problematic
to reduce subloop insertion to two-point function using the same approach.
Solution to this problem can be found in the application of the Feynman
trick to the groups of the propagators each carrying the same integration
momenta of the both loops. Thus, we can reduce crossed subloop insertion
to two-point function. In this paper, we start with the general outline
of the methods proposed in \cite{AA1} and then consider two specific
two-loop topologies, crossed two-loop vertex and box diagrams, and
develop a generalized approach on how to treat crossed subloop insertion
in two-loop calculations. 

\section{Subloop Insertion}

Idea of the subloop insertion was employed in \cite{AA1} for the
case of two-loop topologies. Before considering crossed two-loop triangle
and box topology, let us review general ideas developed in \cite{AA1}
for the triangle and box type subloop insertion. 
\begin{figure}
\begin{centering}
\includegraphics[scale=0.35]{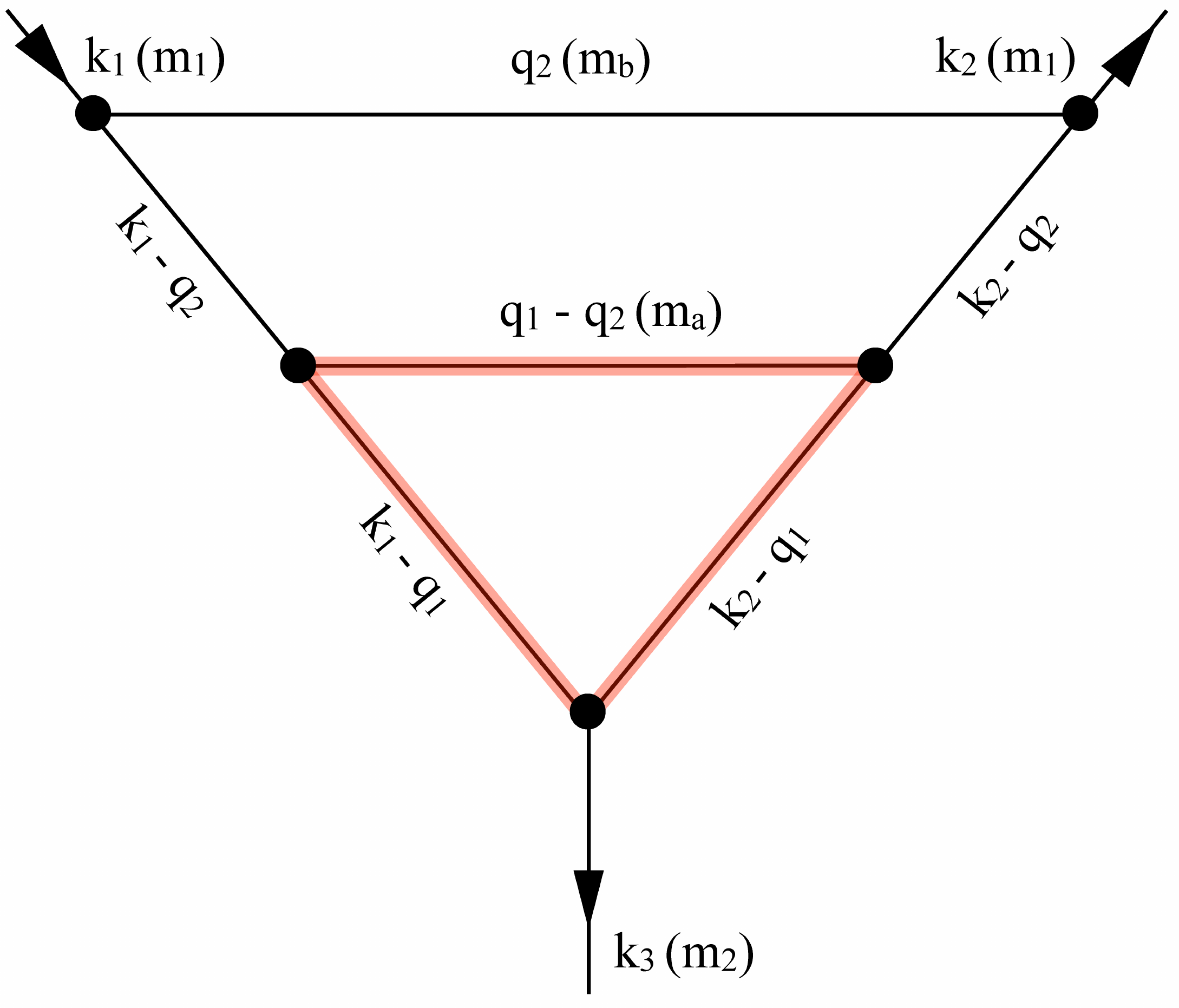}
\par\end{centering}
\caption{Two-loop triangle topology with triangle type insertion. Red and bold
selection corresponds to the triangle subloop insertion.}

\label{fig1}
\end{figure}
Let us start with two-loop triangle topology shown on Fig.(\ref{fig1}).
To simplify our derivations, we consider the case where all the couplings
are set to one, and all the particles are scalars. We will assume
that particles carrying momenta $k_{1}$ and $k_{2}$ are on-shell,
and particle with momentum $k_{3}$ is off-shell. Here, we can write
\begin{align}
I_{\Delta_{1}}= & -\frac{1}{\pi^{4}}\intop\frac{d^{4}q_{1}d^{4}q_{2}}{\left[\left(k_{1}-q_{1}\right)^{2}-m_{1}^{2}\right]\left[\left(k_{2}-q_{1}\right)^{2}-m_{1}^{2}\right]\left[\left(q_{1}-q_{2}\right)^{2}-m_{a}^{2}\right]\left[\left(k_{2}-q_{2}\right)^{2}-m_{1}^{2}\right]}\nonumber \\
\nonumber \\
 & \times\frac{1}{\left[q_{2}^{2}-m_{b}^{2}\right]\left[\left(k_{1}-q_{2}\right)^{2}-m_{1}^{2}\right]}.\label{eq:11}
\end{align}
The first three propagators in Eq.(\ref{eq:11}) belong to the triangle
insertion (red and bold ) in Fig.(\ref{fig1}). Initial step is to
join first two propagators, without momenta of the second loop, using
Feynman trick:
\begin{align}
I_{\Delta_{1}}= & -\frac{1}{\pi^{4}}\intop_{0}^{1}dx\intop\frac{d^{4}q_{1}d^{4}q_{2}}{\left[\left(q_{1}-\left(xk_{1}+\bar{x}k_{2}\right)\right)^{2}-\left(xk_{1}+\bar{x}k_{2}\right)^{2}\right]^{2}\left[\left(q_{1}-q_{2}\right)^{2}-m_{a}^{2}\right]\left[\left(k_{2}-q_{2}\right)^{2}-m_{1}^{2}\right]}\nonumber \\
\nonumber \\
 & \times\frac{1}{\left[q_{2}^{2}-m_{b}^{2}\right]\left[\left(k_{1}-q_{2}\right)^{2}-m_{1}^{2}\right]},\label{eq:12}
\end{align}
where $\bar{x}=1-x.$ Quadratic form in Eq.(\ref{eq:12}) can be removed
if we apply so-called mass shift approach:
\begin{align}
\frac{1}{\left[\left(q_{1}-\left(xk_{1}+\bar{x}k_{2}\right)\right)^{2}-\left(xk_{1}+\bar{x}k_{2}\right)^{2}\right]^{2}} & =\lim_{\lambda\rightarrow0}\frac{\partial}{\partial\lambda}\frac{1}{\left(q_{1}-\left(xk_{1}+\bar{x}k_{2}\right)\right)^{2}-\left(\left(xk_{1}+\bar{x}k_{2}\right)^{2}+\lambda\right)}.\label{eq:13}
\end{align}
After substituting momentum $q_{1}=\tau+xk_{1}+\bar{x}k_{2}$ and
using Eq.(\ref{eq:13}), we can replace the first loop integral in
Eq.(\ref{eq:12}) by two-point Passarino-Veltman function: 
\begin{align}
 & I_{\Delta_{1}}=-\frac{i}{\pi^{2}}\lim_{\lambda\rightarrow0}\frac{\partial}{\partial\lambda}\intop_{0}^{1}dx\intop d^{4}q_{2}\frac{B_{0}\left[\left(q_{2}-xk_{1}-\bar{x}k_{2}\right)^{2},m_{\lambda}^{2},m_{a}^{2}\right]}{\left[\left(k_{2}-q_{2}\right)^{2}-m_{1}^{2}\right]\left[q_{2}^{2}-m_{b}^{2}\right]\left[\left(k_{1}-q_{2}\right)^{2}-m_{1}^{2}\right]}.\label{eq:14}
\end{align}
Here, $m_{\lambda}^{2}=\left(xk_{1}+\bar{x}k_{2}\right)^{2}+\lambda=m_{1}^{2}-x\bar{x}k_{3}^{2}+\lambda$.
The next step is to apply dispersive replacement of the two-point
function:
\begin{align}
 & I_{\Delta_{1}}=\frac{i}{\pi^{3}}\lim_{\lambda\rightarrow0}\frac{\partial}{\partial\lambda}\intop_{0}^{1}dx\intop_{\left(m_{a}+m_{\lambda}\right)^{2}}^{\Lambda^{2}}ds\,\Im B_{0}\left[s,m_{\lambda}^{2},m_{a}^{2}\right]\nonumber \\
\nonumber \\
 & \times\intop\frac{d^{4}q_{2}}{\left[\left(k_{2}-q_{2}\right)^{2}-m_{1}^{2}\right]\left[q_{2}^{2}-m_{b}^{2}\right]\left[\left(k_{1}-q_{2}\right)^{2}-m_{1}^{2}\right]\left[\left(q_{2}-xk_{1}-\bar{x}k_{2}\right)^{2}-s-i\epsilon\right]}.\label{eq:15}
\end{align}
As a result, the denominator $\frac{1}{\left(q_{2}-xk_{1}-\bar{x}k_{2}\right)^{2}-s-i\epsilon}$,
which is coming from dispersive representation of two-point function
in Eq.(\ref{eq:14}), is absorbed into the second loop integration
as an additional propagator. As it was discussed in \cite{AA1}, effective
mass parameter $m_{\lambda}^{2}$ could become negative (if $k_{3}^{2}>\frac{m_{1}^{2}+\lambda}{x\bar{x}}$)
and that requires a different treatment of dispersive integral. This
is discussed in details in {[}1{]} and it is straightforward to implement
the case where $m_{\lambda}^{2}<0$. To avoid lengthy expressions,
we will assume a condition where $m_{\lambda}^{2}$ is positive. In
the next step, in Eq.(\ref{eq:15}), we can apply Feynman trick to
the first three propagators, and after using mass shift approach we
can write the final two-loop result in two-point function basis:
\begin{align}
 & I_{\Delta_{1}}=-\frac{1}{\pi}\lim_{\left\{ \lambda,\delta\right\} \rightarrow0}\frac{\partial^{3}}{\partial\lambda\partial\delta^{2}}\intop_{0}^{1}dxdy\intop_{0}^{1-y}dz\intop_{\left(m_{a}+m_{\lambda}\right)^{2}}^{\Lambda^{2}}ds\,\Im B_{0}\left[s,m_{\lambda}^{2},m_{a}^{2}\right]\nonumber \\
\nonumber \\
 & \times B_{0}\left[\left(\left(y-\bar{x}\right)k_{2}+\left(z-x\right)k_{1}\right)^{2},m_{\delta}^{2},s\right].\label{eq:16}
\end{align}
Effective mass $m_{\delta}^{2}$ is defined as follows: $m_{\delta}^{2}=\left(\bar{y}-z\right)m_{b}^{2}+\left(y+z\right)^{2}m_{1}^{2}-yzk_{3}^{2}+\delta$.
Integration cutoff $\Lambda^{2}$ is introduced in order to keep the
integration finite. After differentiation with respect to $\lambda$
and $\delta$, the dependence on cutoff and regularization parameters
in Eq,(\ref{eq:16}) will cancel. Using the approach outlined in derivation
of Eq.(\ref{eq:16}), we can express two-loop triangle graph with
arbitrary tensorial rank in two-point function basis analytically,
and later perform integration and differentiation numerically. If
there are ultraviolet divergences, they can be addressed by employing
the subloop subtraction in a given renormalization scheme. In this
case, the dispersive integral in Eq.(\ref{eq:16}) will have a singly-
or doubly-subtracted structure. The second loop renormalization can
be achieved by adding second-order counter terms, also computed using
dispersive representation. The same ideas can be applied to the box
subloop. 
\begin{figure}
\begin{centering}
\includegraphics[scale=0.35]{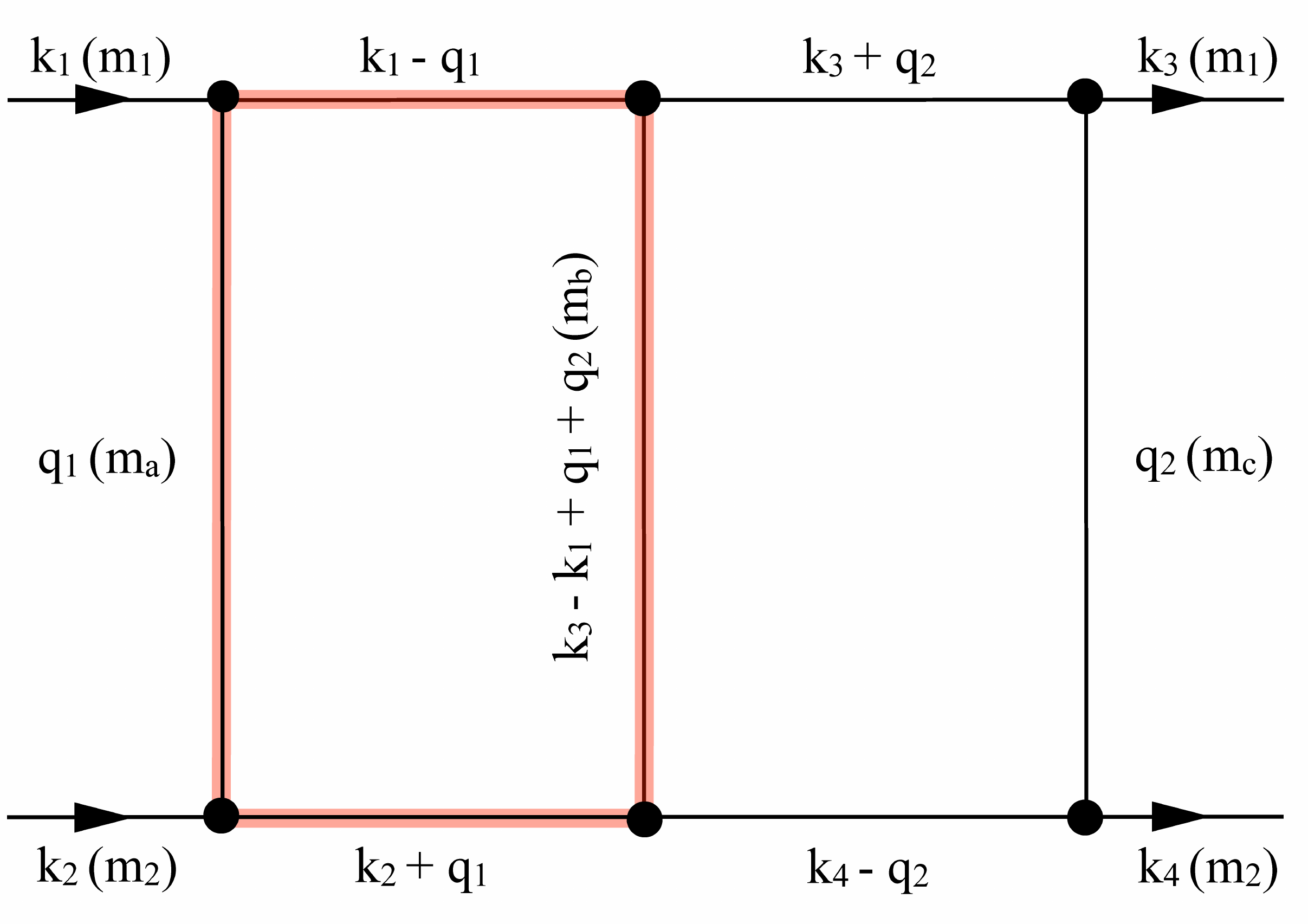}
\par\end{centering}
\caption{Two-loop box topology (double box) with box type insertion. Red and
bold selection corresponds to the box subloop insertion.}

\label{fig2}
\end{figure}
If we consider the diagram on Fig.(\ref{fig2}), we can write the
following:
\begin{align}
I_{\boxempty_{1}} & =-\frac{1}{\pi^{4}}\intop\frac{d^{4}q_{1}d^{4}q_{2}}{\left[q_{1}^{2}-m_{a}^{2}\right]\left[\left(k_{1}-q_{1}\right)^{2}-m_{1}^{2}\right]\left[\left(k_{2}+q_{1}\right)^{2}-m_{2}^{2}\right]\left[\left(k_{3}-k_{1}+q_{1}+q_{2}\right)^{2}-m_{b}^{2}\right]}\nonumber \\
\nonumber \\
 & \times\frac{1}{\left[q_{2}^{2}-m_{c}^{2}\right]\left[\left(k_{4}-q_{2}\right)^{2}-m_{2}^{2}\right]\left[\left(k_{3}+q_{2}\right)^{2}-m_{1}^{2}\right]}.\label{eq:17}
\end{align}
After joining the first three propagators, shifting momentum $q_{1}=\tau-q_{2}-k_{3}+k_{1}$,
and expressing two-point function by dispersive integral, we can write:
\begin{align}
I_{\boxempty_{1}} & =\frac{i}{\pi^{3}}\lim_{\lambda\rightarrow0}\frac{\partial^{2}}{\partial\lambda^{2}}\intop_{0}^{1}dx\intop_{0}^{1-x}dy\intop_{\left(m_{b}+m_{\lambda}\right)^{2}}^{\Lambda^{2}}ds\,\Im B_{0}\left[s,m_{b}^{2},m_{\lambda}^{2}\right]\nonumber \\
\nonumber \\
 & \times\intop\frac{d^{4}q_{2}}{\left[q_{2}^{2}-m_{c}^{2}\right]\left[\left(k_{4}-q_{2}\right)^{2}-m_{2}^{2}\right]\left[\left(k_{3}+q_{2}\right)^{2}-m_{1}^{2}\right]\left[\left(q_{2}+k_{3}-xk_{2}-k_{1}\bar{y}\right)^{2}-s-i\epsilon\right]}.\label{eq:18}
\end{align}
Here, effective mass $m_{\lambda}$defined as $m_{\lambda}^{2}=m_{a}^{2}\left(\bar{x}-y\right)+x^{2}m_{2}^{2}+y^{2}m_{1}^{2}-2xy\left(k_{1}k_{2}\right)+\lambda$.
In the same way as before, after joining the first three propagators
in the second loop integral, we get the following two-loop box result:
\begin{align}
I_{\boxempty_{1}} & =-\frac{1}{\pi}\lim_{\lambda\rightarrow0}\frac{\partial^{4}}{\partial\lambda^{2}\partial\delta^{2}}\intop_{0}^{1}dxdz\intop_{0}^{1-x}dy\intop_{0}^{1-z}d\omega\intop_{\left(m_{b}+m_{\lambda}\right)^{2}}^{\Lambda^{2}}ds\,\Im B_{0}\left[s,m_{b}^{2},m_{\lambda}^{2}\right]\nonumber \\
\nonumber \\
 & \times B_{0}\left[\left(\omega k_{4}+\bar{z}k_{3}-xk_{2}-\bar{y}k_{1}\right)^{2},m_{\delta}^{2},s\right].\label{eq:19}
\end{align}
Effective mass $m_{\delta}$ has the following structure: $m_{\delta}^{2}=m_{c}^{2}\left(\bar{z}-\omega\right)+m_{1}^{2}z^{2}+m_{2}^{2}\omega^{2}-2z\omega\left(k_{3}k_{4}\right)+\delta$.
Results in both Eq.(\ref{eq:16}) and (\ref{eq:19}) are in compact
form and can be implemented in computer algebra-based packages. Endpoint
for two-loop calculations would be numerical evaluation of derivatives
with respect to mass shift parameters, Feynman and dispersion integrals.
Both examples which we have considered here assume that it is possible
to join all propagator except one in the subloop insertion. All joined
propagators should carry integration momentum of the subloop insertion
only. As a result, subloop integral can be replaced by the two-point
function. However, in the case of crossed two-loop topologies, it
is not possible to achieve the same using the outlined approach directly.
Crossed two-loop topologies will have box type insertion subloop with
more than one propagator carrying integration momenta of the first
and second loop. In the next section, we will consider two examples
of crossed two-loop topology, from which we develop an approach allowing
us to express subloop insertion in the two-point function basis, allowing
to write final expressions in a compact form suitable for the numerical
evaluations. 

\section{Crossed Topology Subloop Insertion}

\subsection{Two-Loop Crossed Triangle}

\begin{figure}
\begin{centering}
\includegraphics[scale=0.35]{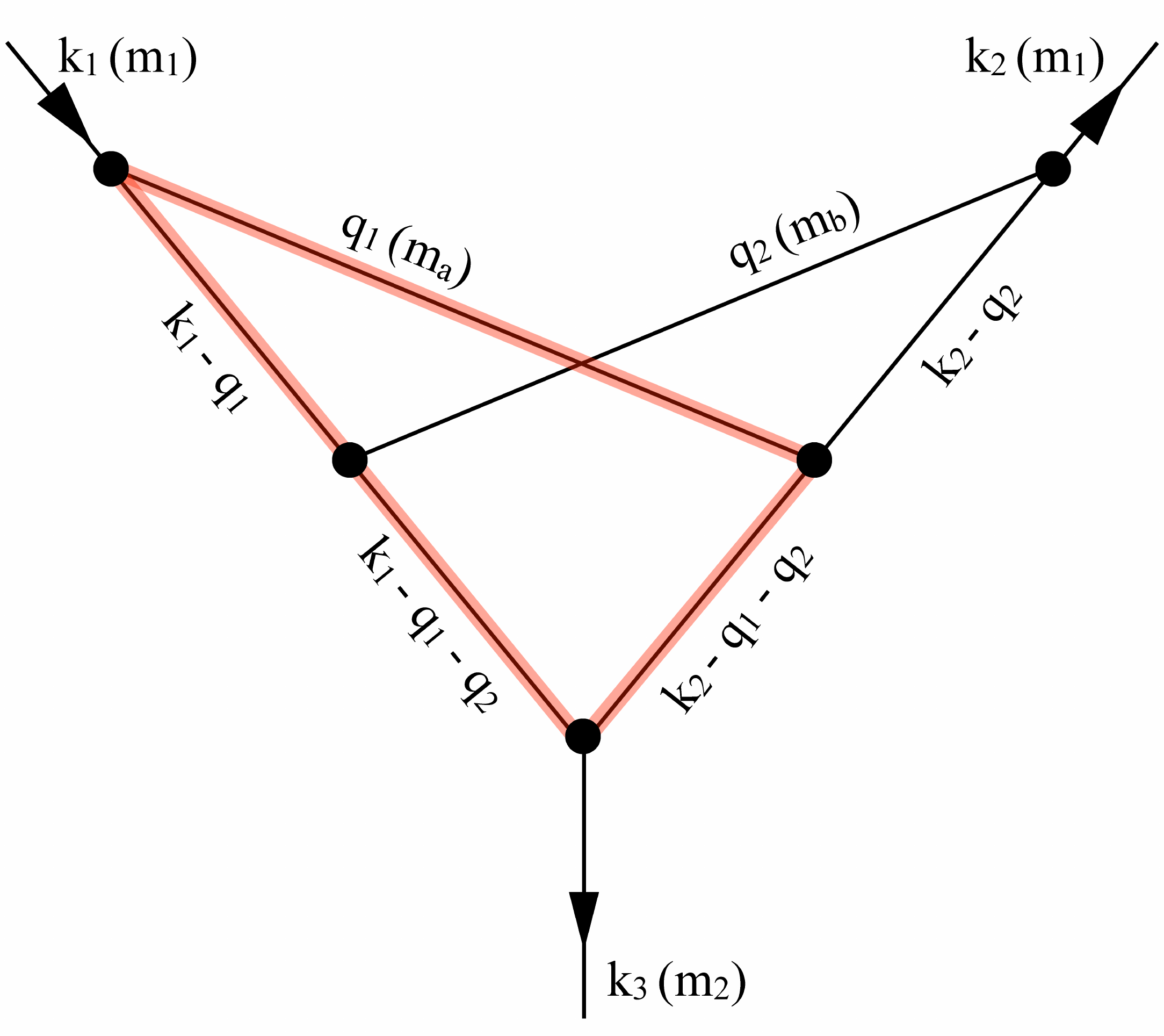}
\par\end{centering}
\caption{Crossed two-loop vertex topology.}

\label{fig3}
\end{figure}

Let us start with the two-loop topology shown on Fig.(\ref{fig3}).
Particles with momenta $q_{1}$, $q_{2}$ and $k_{3}$ have masses
$m_{a}$, $m_{b}$ and $m_{3}$, respectively. All other lines on
graph from Fig.(\ref{fig3}) have the mass $m_{1}$. We will employ
the same idea of dispersive insertion as before, with the final two-loop
result will be given completely in two-point Passarino-Veltman function
basis. 
According to the momenta distribution on Fig.(\ref{fig3}), we can
write the following:
\begin{align}
I_{\Delta_{2}} & =-\frac{1}{\pi^{4}}\intop\frac{d^{4}q_{1}d^{4}q_{2}}{\left[q_{1}^{2}-m_{a}^{2}\right]\left[\left(k_{1}-q_{1}\right)^{2}-m_{1}^{2}\right]\left[\left(k_{1}-q_{1}-q_{2}\right)^{2}-m_{1}^{2}\right]\left[\left(k_{2}-q_{1}-q_{2}\right)^{2}-m_{1}^{2}\right]}\nonumber \\
\nonumber \\
 & \times\frac{1}{\left[q_{2}^{2}-m_{b}^{2}\right]\left[\left(k_{2}-q_{2}\right)^{2}-m_{1}^{2}\right]}.\label{eq:1}
\end{align}
As noted before, the subloop insertion in Fig.(\ref{fig3}) has only
two propagators without the momentum of the second loop. In order
to reduce subloop into two-point function, we will join the first
and second propagators and then third and fourth. In this case we
get the following: 
\begin{align}
I_{\Delta_{2}} & =-\frac{1}{\pi^{4}}\lim_{\left\{ \xi,\,\lambda\right\} \rightarrow0}\frac{\partial^{2}}{\partial\xi\partial\lambda}\intop_{0}^{1}dxdy\intop\frac{d^{4}q_{2}}{\left[q_{2}^{2}-m_{b}^{2}\right]\left[\left(k_{2}-q_{2}\right)^{2}-m_{1}^{2}\right]}\nonumber \\
\nonumber \\
 & \times\intop\frac{d^{4}q_{1}}{\left[\left(q_{1}-k_{1}x\right)^{2}-m_{\xi}^{2}\right]\left[\left(q_{1}+q_{2}-\bar{y}k_{2}-yk_{1}\right)^{2}-m_{\lambda}^{2}\right]}.\label{eq:2}
\end{align}
Here, masses $m_{\xi}^{2}$ and $m_{\lambda}^{2}$ are defined as
$m_{\xi}^{2}=m_{a}^{2}\bar{x}+m_{1}^{2}x^{2}+\xi$ and $m_{\lambda}^{2}=m_{1}^{2}-\bar{y}yk_{3}^{2}+\lambda$.
Thus, the loop integral over $q_{1}$, after replacing $q_{1}=\tau+k_{1}x$,
now can be written as a two-point function:
\begin{align}
I_{\Delta_{2}} & =-\frac{i}{\pi^{2}}\lim_{\left\{ \xi,\,\lambda\right\} \rightarrow0}\frac{\partial^{2}}{\partial\xi\partial\lambda}\intop_{0}^{1}dxdy\intop d^{4}q_{2}\frac{B_{0}\left[\left(q_{2}-\bar{y}k_{2}+k_{1}\left(x-y\right)\right)^{2},m_{\xi}^{2},m_{\lambda}^{2}\right]}{\left[q_{2}^{2}-m_{b}^{2}\right]\left[\left(k_{2}-q_{2}\right)^{2}-m_{1}^{2}\right]}.\label{eq:3}
\end{align}
After replacing the two-point function by it's dispersive representation,
we can address evaluation of the second loop integral:
\begin{align}
 & I_{\Delta_{2}}=\frac{i}{\pi^{3}}\lim_{\left\{ \xi,\,\lambda\right\} \rightarrow0}\frac{\partial^{2}}{\partial\xi\partial\lambda}\intop_{0}^{1}dxdy\intop_{\left(m_{\xi}+m_{\lambda}\right)^{2}}^{\Lambda^{2}}ds\,\Im B_{0}\left[s,m_{\xi}^{2},m_{\lambda}^{2}\right]\nonumber \\
\nonumber \\
 & \times\intop\frac{d^{4}q_{2}}{\left[q_{2}^{2}-m_{b}^{2}\right]\left[\left(k_{2}-q_{2}\right)^{2}-m_{1}^{2}\right]\left[\left(q_{2}-\bar{y}k_{2}+k_{1}\left(x-y\right)\right)^{2}-s-i\epsilon\right]}.\label{eq:4}
\end{align}
Now, after joining the first two propagators in Eq.(\ref{eq:4}) and
introducing mass shift parameter $\phi$, the final result for the
graph on Fig.(\ref{fig3}) can be expressed in terms of the product
of two two-point functions, which are later integrated and then numerically
differentiated:
\begin{align}
I_{\Delta_{2}} & =-\frac{1}{\pi}\lim_{\left\{ \xi,\,\lambda,\,\phi\right\} \rightarrow0}\frac{\partial^{3}}{\partial\xi\partial\lambda\partial\phi}\intop_{0}^{1}dxdydz\intop_{\left(m_{\xi}+m_{\lambda}\right)^{2}}^{\Lambda^{2}}ds\,\Im B_{0}\left[s,m_{\xi}^{2},m_{\lambda}^{2}\right]\nonumber \\
\nonumber \\
 & \times B_{0}\left[\left(\left(z-\bar{y}\right)k_{2}+k_{1}\left(x-y\right)\right)^{2},m_{\phi}^{2},s\right].\label{eq:5}
\end{align}
Here, mass $m_{\phi}^{2}$ has the following structure: $m_{\phi}^{2}=m_{b}^{2}\bar{z}+m_{1}^{2}z^{2}+\phi$.
Eqs.(\ref{eq:16}) and (\ref{eq:5}) have the same dimension of multidimensional
integration and the same overall order of the differentiation with
respect to the mass shift parameters. In addition to that, in the
derivation of Eq.(\ref{eq:5}), we had to introduce one extra mass
shift parameter. It is obvious that in the crossed-type subloop insertion
we would have to deal with four propagators. General structure of
the insertion would have two groups of propagators with the similar
momenta in each group. That allows us to join propagators in two groups
separately and effectively reduce an entire insertion to the two-point
function. The same approach could be implemented for the two-loop
crossed box topology discussed in the next subsection.

\subsection{Two-Loop Crossed Box}

Let us start by writing a general expression for the two-loop integral
for a crossed two-loop box topology shown on Fig.(\ref{fig4}):
\begin{figure}
\begin{centering}
\includegraphics[scale=0.35]{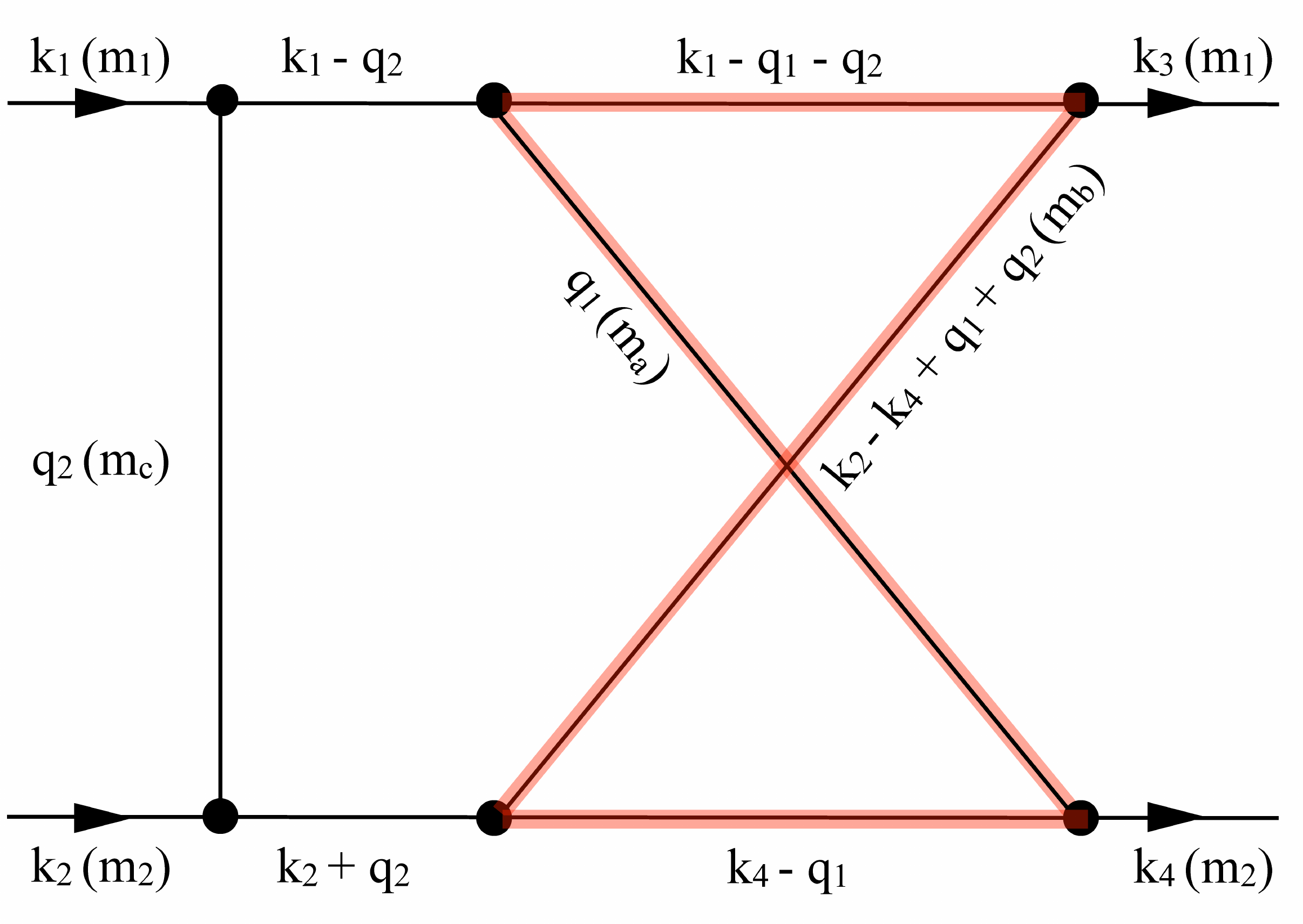}
\par\end{centering}
\caption{Crossed two-loop box topology.}

\label{fig4}
\end{figure}
\begin{align}
I_{\boxempty_{2}} & =-\frac{1}{\pi^{4}}\intop\frac{d^{4}q_{1}d^{4}q_{2}}{\left[q_{1}^{2}-m_{a}^{2}\right]\left[\left(k_{4}-q_{1}\right)^{2}-m_{2}^{2}\right]\left[\left(k_{2}-k_{4}+q_{1}+q_{2}\right)^{2}-m_{b}^{2}\right]\left[\left(k_{1}-q_{1}-q_{2}\right)^{2}-m_{1}^{2}\right]}\nonumber \\
\nonumber \\
 & \times\frac{1}{\left[q_{2}^{2}-m_{c}^{2}\right]\left[\left(k_{1}-q_{2}\right)^{2}-m_{1}^{2}\right]\left[\left(k_{2}+q_{2}\right)^{2}-m_{2}^{2}\right]}.\label{eq:6}
\end{align}
After joining the first and the second, and then the third and the
fourth propagators, we can write, with the help of two mass shift
parameters $\xi$ and $\lambda$, the following:
\begin{align}
I_{\boxempty_{2}} & =-\frac{i}{\pi^{2}}\lim_{\left\{ \xi,\,\lambda\right\} \rightarrow0}\frac{\partial^{2}}{\partial\xi\partial\lambda}\intop_{0}^{1}dxdy\intop d^{4}q_{2}\frac{B_{0}\left[\left(q_{2}-xk_{4}+yk_{3}-k_{1}\right)^{2},m_{\xi}^{2},m_{\lambda}^{2}\right]}{\left[q_{2}^{2}-m_{c}^{2}\right]\left[\left(k_{1}-q_{2}\right)^{2}-m_{1}^{2}\right]\left[\left(k_{2}+q_{2}\right)^{2}-m_{2}^{2}\right]},\label{eq:7}
\end{align}
where $m_{\xi}^{2}=\bar{x}m_{a}^{2}+x^{2}m_{2}^{2}+\xi$ and $m_{\lambda}^{2}=\bar{y}^{2}m_{1}^{2}+ym_{b}^{2}+\lambda$.
Replacing two-point function in Eq.(\ref{eq:7}) by a dispersion integral,
we arrive to: 
\begin{align}
I_{\boxempty_{2}} & =\frac{i}{\pi^{3}}\lim_{\left\{ \xi,\,\lambda\right\} \rightarrow0}\frac{\partial^{2}}{\partial\xi\partial\lambda}\intop_{0}^{1}dxdy\intop_{\left(m_{\xi}+m_{\lambda}\right)^{2}}^{\Lambda^{2}}ds\,\Im B_{0}\left[s,m_{\xi}^{2},m_{\lambda}^{2}\right]\nonumber \\
\label{eq:8}\\
 & \times\intop\frac{d^{4}q_{2}}{\left[q_{2}^{2}-m_{c}^{2}\right]\left[\left(k_{1}-q_{2}\right)^{2}-m_{1}^{2}\right]\left[\left(k_{2}+q_{2}\right)^{2}-m_{2}^{2}\right]\left[\left(q_{2}-xk_{4}+yk_{3}-k_{1}\right)^{2}-s-i\epsilon\right]}.\nonumber 
\end{align}
The second loop integration is done after joining first three propagators
in the integral over $q_{2}$ and introducing third mass shift parameter
$\delta$:
\begin{align}
I_{\boxempty_{2}} & =-\frac{1}{\pi}\lim_{\left\{ \xi,\,\lambda\right\} \rightarrow0}\frac{\partial^{4}}{\partial\xi\partial\lambda\partial\delta^{2}}\intop_{0}^{1}dxdydz\intop_{0}^{1-z}dw\intop_{\left(m_{\xi}+m_{\lambda}\right)^{2}}^{\Lambda^{2}}ds\,\Im B_{0}\left[s,m_{\xi}^{2},m_{\lambda}^{2}\right]\nonumber \\
\nonumber \\
 & \times B_{0}\left[\left(\bar{z}k_{1}+\omega k_{2}+xk_{4}-yk_{3}\right)^{2},m_{\delta}^{2},s\right],\label{eq:9}
\end{align}
where $m_{\delta}^{2}=\left(\bar{z}-\omega\right)m_{c}^{2}+\left(\omega k_{2}-zk_{1}\right)^{2}+\delta$.
The crossed two-loop box also acquires an additional mass shift parameter,
while a degree of multidimensional integration and general order of
differentiation remains the same as in Eq.(\ref{eq:19}). It is evident
that in both cases reflected on Fig.(\ref{fig3}) and Fig.(\ref{fig4}),
we are dealing with the box-type insertion. In the cases without a
crossed topology, external legs of an insertion would carry momentum
of the second loop sequentially. For example, if we have external
momenta labeled as $p_{1},$ $p_{2}$, $p_{3}$ and $p_{4}$, the
second loop momentum shows up in the combinations of momenta $\left\{ p_{1},p_{2}\right\} $,
$\left\{ p_{2},p_{3}\right\} $, $\left\{ p_{3},p_{4}\right\} $ and
$\left\{ p_{1},p_{4}\right\} $. This will allow us to join three
propagators without momentum of the second loop, and arrive to the
results for the box-type insertion outlined in {[}1{]}. The crossed-type
box insertions will have the second loop momentum appear in the combinations
of external momenta such as $\left\{ p_{1},p_{3}\right\} $ and $\left\{ p_{2},p_{4}\right\} $.
As a result, we would have to apply Feynman trick to two groups of
propagators. At this point, we will consider a general case of the
crossed box type subloop insertion. For the crossed box subloop, shown
on Fig.(\ref{fig5}), we will assume that external momenta $p_{2}$
and $p_{4}$ would depend on the momentum of the second loop.
\begin{figure}
\begin{centering}
\includegraphics[scale=0.35]{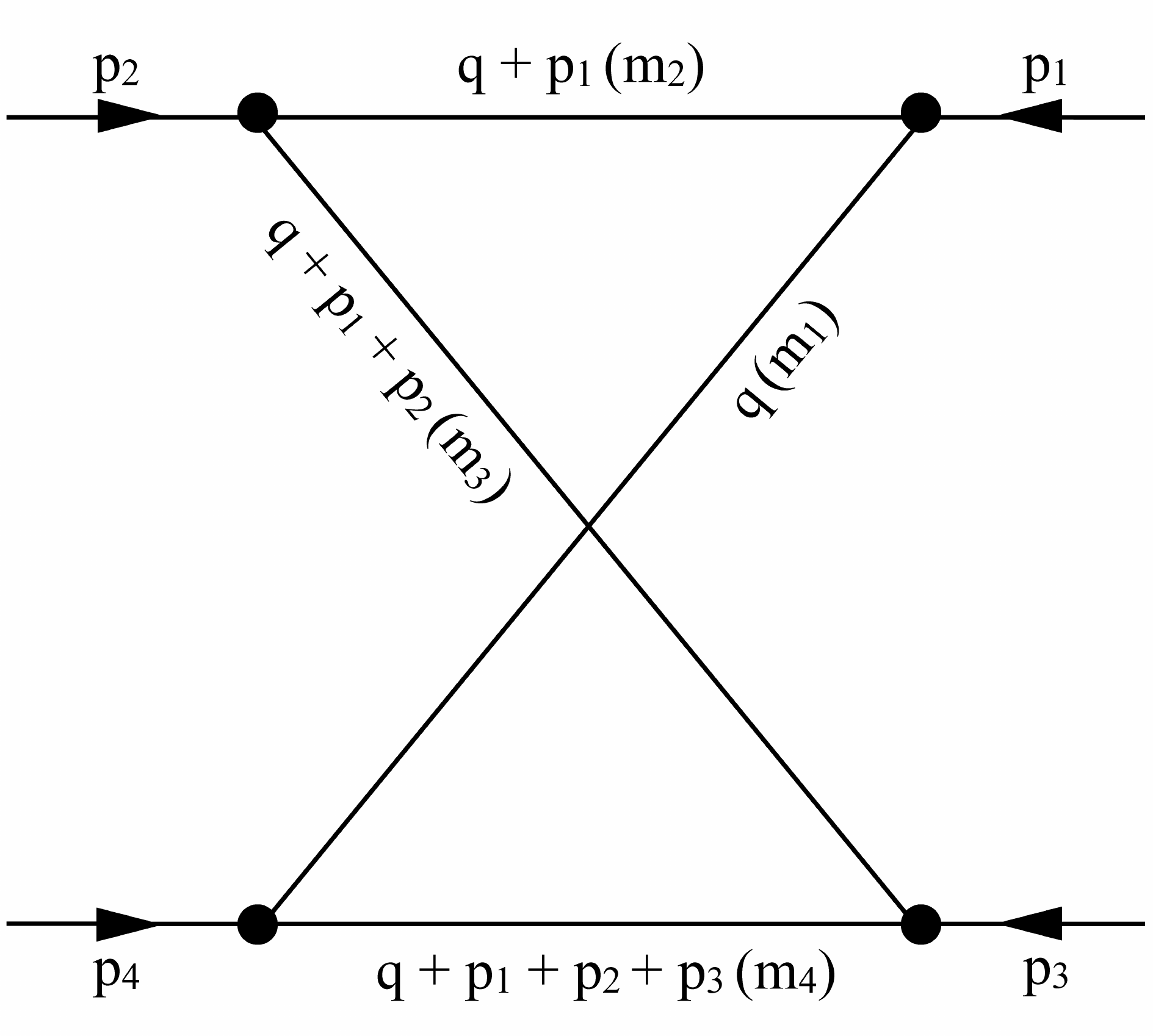}
\par\end{centering}
\caption{General crossed box subloop.}

\label{fig5}
\end{figure}
In this case, we can write:
\begin{align}
I_{\boxtimes} & =\frac{1}{i\pi^{2}}\intop\frac{d^{4}q}{\left[q^{2}-m_{1}^{2}\right]\left[\left(q+p_{1}\right)^{2}-m_{2}^{2}\right]\left[\left(q+p_{1}+p_{2}\right)^{2}-m_{3}^{2}\right]\left[\left(q+p_{1}+p_{2}+p_{3}\right)^{2}-m_{4}^{2}\right]}.\label{eq:10}
\end{align}
As it was considered in the previous examples, we will join propagators
in Eq.(\ref{eq:10}) in two groups: the first and the second, and
then the third and the fourth. After introducing two mass shift parameters,
and shifting momentum of integration $q=\tau-xp_{1},$ we can rewrite
Eq.(\ref{eq:10}) in the two-point function basis:
\begin{align}
I_{\boxtimes} & =\lim_{\left\{ \xi,\lambda\right\} \rightarrow0}\frac{\partial^{2}}{\partial\xi\partial\lambda}\intop_{0}^{1}dxdy\,B_{0}\left[\left(\bar{x}p_{1}+p_{2}+yp_{3}\right)^{2},m_{\xi}^{2},m_{\lambda}^{2}\right].\label{eq:11-1}
\end{align}
Here, $m_{\xi}^{2}=\bar{x}m_{1}^{2}+xm_{2}^{2}-x\bar{x}p_{1}^{2}+\xi$
and $m_{\lambda}^{2}=\bar{y}m_{3}^{2}+ym_{4}^{2}+p_{1}^{2}+yp_{3}^{2}+\lambda$.
Replacing the two-point function in Eq.(\ref{eq:11-1}) by a dispersive
representation, we arrive to the following result:
\begin{align}
I_{\boxtimes} & =-\frac{1}{\pi}\lim_{\left\{ \xi,\lambda\right\} \rightarrow0}\frac{\partial^{2}}{\partial\xi\partial\lambda}\intop_{0}^{1}dxdy\intop_{\left(m_{\xi}+m_{\lambda}\right)^{2}}^{\Lambda^{2}}ds\,\frac{\Im B_{0}\left[s,m_{\xi}^{2},m_{\lambda}^{2}\right]}{\left(\bar{x}p_{1}+p_{2}+yp_{3}\right)^{2}-s-i\epsilon}.\label{eq:12-1}
\end{align}
The Eq.(\ref{eq:12-1}) suggests that in general, if we encounter
crossed box subloop insertion, we can replace it by the effective
four-particle coupling:
\begin{align}
\Gamma_{\boxtimes} & =\hat{\boldsymbol{D}}\left[\frac{\Im B_{0}\left[s,m_{\xi}^{2},m_{\lambda}^{2}\right]}{\left(\bar{x}p_{1}+p_{2}+yp_{3}\right)^{2}-s-i\epsilon}\right],\label{eq:13-1}
\end{align}
with operator $\hat{\boldsymbol{D}}$ is defined as $\hat{\boldsymbol{D}}=\lim_{\left\{ \xi,\lambda\right\} \rightarrow0}\frac{\partial^{2}}{\partial\xi\partial\lambda}\intop_{0}^{1}dxdy\intop_{\left(m_{\xi}+m_{\lambda}\right)^{2}}^{\Lambda^{2}}ds....$
Using Eq.(\ref{eq:13-1}), we can perform the second loop integration
in the two-point function basis, and at the end evaluate derivatives
and integrals numerically. In general, it is straightforward to extend
this approach to the cases with tensor-type numerator which was considered
in details in \cite{AA1}.

\section{Conclusion}

In this work, we have addressed a specific type of the crossed topologies
arising in the two-loop triangle and box graphs. Based on the examples
outlined in the paper, we have developed an approach where crossed
subloop insertion can be replaced by two-point function basis. Later,
the two-point function can be represented by a dispersive integral
and an arising propagator-like term can be moved to the second loop
integration. The second loop integration can also be reduced into
the two-point function representation. As a result, we can express
the two-loop matrix elements analytically in a rather compact form,
and the scalar integration over Feynman parameter space and dispersive
integration and differentiation with respect to mass-shift parameters
can be carried out numerically at the last stage. This way, we can
address the problem of the much-needed complete electroweak two-loop
calculation by automatization of the entire process. 
\begin{acknowledgments}
Author is grateful to S. Barkanova for stimulating and inspiring discussions.
This work was supported by National Science and Engineering Research
Council (NSERC) of Canada. 
\end{acknowledgments}

\end{document}